\documentclass[12pt,a4paper,oneside,onecolumn,final]{article}
\usepackage[cp1250]{inputenc}
\usepackage[english]{babel}
\usepackage{amsmath}
\usepackage{amssymb}
\usepackage{graphicx}

\usepackage{setspace}
\usepackage[a4paper,left=2cm,right=2cm,top=2cm,bottom=2cm]{geometry}
\usepackage[square,comma,numbers,compress]{natbib}

\begin{document}
\begin{flushleft}\textit{%
Journal of Experimental and Theoretical Physics, 2009, Vol.~109, No.~3, pp.~393--407.$^*$ \copyright~Pleiades Publishing, Inc., 2009.}
\end{flushleft}
\bigskip
\begin{center}
\begin{Large}
\textbf{The Filling of Neutron Star Magnetospheres with Plasma:
Dynamics of the Motion of Electrons and Positrons}
\end{Large}
\\ \bigskip
\textrm{Ya. N. Istomin}$^{\mathrm{a},\dag}$
\textrm{ and }
\textrm{D. N. Sobyanin}$^{\mathrm{b},\ddag}$
\\ \bigskip
$^\mathrm{a}$\textit{Lebedev Physical Institute, Russian Academy of Sciences,\\
Leninskii pr. 53, Moscow, 119991 Russia}
\\ \medskip
$^\mathrm{b}$\textit{Moscow Institute of Physics and Technology (State University),\\
Institutskii per. 9, Dolgoprudnyi, Moscow oblast, 141700 Russia}
\\ \bigskip
Received September 15, 2008
\\ \bigskip
\begin{large}
\textsf{Abstract}
\end{large}
\end{center}
\vspace{-1em}
\par\textrm{We consider the motion of charged particles in the vacuum magnetospheres of rotating neutron stars with a strong surface magnetic field, $B\gtrsim10^{12}$~G. The electrons and positrons falling into the magnetosphere or produced in it are shown to be captured by the force-free surface~$\mathbf{E}\cdot\mathbf{B}=0$. Using the Dirac–Lorentz equation, we investigate the dynamics of particle capture and subsequent motion near the force-free surface. The particle energy far from the force-free surface has been found to be determined by the balance between the power of the forces of an accelerating electric field and the intensity of curvature radiation. When captured, the particles perform adiabatic oscillations along the magnetic field lines and simultaneously move along the force-free surface. We have found the oscillation parameters and trajectories of the captured particles. We have calculated the characteristic capture times and energy losses of the particles through the emission of both bremsstrahlung and curvature photons by them. The capture of particles is shown to lead to a monotonic increase in the thickness of the layer of charged plasma accumulating near the force-free surface. The time it takes for a vacuum magnetosphere to be filled with plasma has been estimated.}
\vfill\noindent
\bigskip\\
\rule{\textwidth}{0.1pt}
\noindent
\begin{spacing}{1}
{\footnotesize\noindent
$^\dag$E-mail: \texttt{istomin@lpi.ru}\\
$^\ddag$E-mail: \texttt{sobyanin@gmail.com}\\
$^*$Original Russian Text \copyright Ya.N. Istomin, D.N. Sobyanin, 2009, published in Zhurnal \'{E}ksperimental'noi i Teoreticheskoi Fiziki, 2009, Vol.~136, No.~3, pp.~458--475.}
\end{spacing}
\newpage
\section{Introduction}
In the physics of radio pulsars, the stationary structure of radio pulsar magnetospheres has been investigated in considerable detail. We mean not the vacuum magnetosphere but a magnetosphere filled with a dense electron–positron plasma. This is because the radio emission generated in the magnetosphere by a charged particle flux requires the existence of a dense plasma production mechanism for its explanation. This mechanism suggested by Sturrock \citep{Sturrock1971} and developed significantly by Ruderman and Sutherland \citep{RudermanSutherland1975} is that an electron–positron pair can be efficiently produced by a gamma-ray photon with an energy higher than twice the electron rest mass in a strong magnetic field near the surface of a neutron star that is a radio pulsar, $B\approx 10^{12}$~G \citep{Klepikov1954,Erber1966}. In turn, energetic gamma-ray photons are emitted by electrons and positrons as they move in the magnetosphere along magnetic field lines with a significant curvature. These photons are called curvature ones. The photon emission and absorption in a magnetic field accompanied by the production of electron–positron pairs allow a theory of stationary plasma generation in the magnetosphere of a magnetized rotating neutron star to be constructed. However, from observations of stationarily operating radio pulsars, it is rather difficult to understand what the generation mechanism of radio emission in their magnetospheres is and how the plasma is produced. Tracing the dynamics of emission development at various frequencies would be very important for understanding the physical processes in the magnetospheres of radio pulsars~\citep{GurevichIstomin2007}. In addition, many observations of nonstationarily operating radio pulsars have appeared recently. These primarily include the so-called intermittent radio pulsars whose radio emission is observed only during a certain time interval that exceeds considerably the spin period of the star. Thus, for example, the pulsar PSR B1931+24 “operates” for 5–-10 days and
then remains silent for 20–-25 days \citep{KramerEtal2006}, while the pulsar PSR J1832+0029 “operates” for about 300 days and remains silent for about 700 days (see, e.g., the review~\citep{Kramer2008}). The spindown, i.e., the energy losses, was measured to be significantly different in the periods of operation and silence. Since the radio emission power accounts for a negligible fraction of the total rotational energy losses of a neutron star, it would be natural to assume that the silence is related to the termination of plasma generation in the magnetosphere. The mechanism of losses related to the emission of a magnetodipole wave in a vacuum (silence) can then be distinguished from the mechanism of losses related to the emission of a pulsar wind—--an electron–positron
plasma flow from the magnetosphere (operation).

Apart from intermittent pulsars, the group of so-called nulling pulsars has long been known, in which no radio emission is observed for some time but not so regularly as in intermittent pulsars and for which the difference in spindown has not yet been measured.
The nulling fraction exceeds~40\% for seven of the 23 pulsars investigated in~\citep{WangEtal2007} and reaches~95\% for PSR J1502-5653 and PSR J1717-4054.

Yet another group of nonstationary radio sources has been observed recently—--rotating radio transients (RRATs), i.e., sporadically flaring radio sources in which the flare repetition phase is retained and for which the corresponding periods typical of normal
radio pulsars have been measured~\citep{McLaughlinEtal2006}. There is no doubt that these are also rotating neutron stars. Nevertheless, the nature of their activity is completely unclear, as suggested by the existence of many dissimilar models (see, e.g., \citep{ZhuXu2006,ZhangEtal2007,Li2006,LomiashviliEtal2007,OuyedEtal2008}).

In our view, all of the described sources show nonstationary plasma generation processes in the magnetospheres of neutron stars. Therefore, it is important to understand how the magnetosphere of a rotating magnetized neutron star is filled with plasma, which triggers the operation of radio pulsars~\citep{GurevichIstomin2007}. This paper is devoted to an initial investigation of the “ignition” of neutron star magnetospheres---the dynamics of magnetosphere filling with electrons and positrons produced in the magnetosphere. In the Section~2, we give the electromagnetic fields of the inner vacuum magnetosphere and describe the force-free surface where the electric field along the magnetic field becomes zero. Sections 3–-7 are devoted to the dynamics of charged particle motion near the force-free surface and their capture onto this surface. In the Section~8, we discuss the dynamics of plasma accumulation in the inner magnetosphere and estimate the characteristic ignition time.

\section{Structure of the vacuum magnetosphere}

It will be convenient for us to consider the electromagnetic field around the neutron star in spherical coordinates~$r$, $\theta$, and~$\varphi$. We will take the rotation axis of the star specified by the direction of its angular velocity vector~$\mathbf{\Omega}$ as the polar axis. Here, $r$~is the distance from the stellar center to a given point, $\theta$~is the polar angle measured from the rotation axis, and $\varphi$~is the azimuthal angle. The electromagnetic field outside the neutron star was found by Deutsch~\citep{Deutsch1955} to have the following form: the magnetic field
\begin{equation}
\label{deutschMagneticField}
\begin{aligned}
B_r&=\frac{2m}{r^3}\,\Bigl[\cos\theta\cos\theta_m+\sin\theta\sin\theta_m\cos(\varphi-\varphi_m)\Bigr],\\
B_\theta&=\frac{m}{r^3}\,\Bigl[\sin\theta\cos\theta_m-\cos\theta\sin\theta_m\cos(\varphi-\varphi_m)\Bigr],\\
B_\varphi&=\frac{m}{r^3}\,\sin\theta_m\sin(\varphi-\varphi_m);
\end{aligned}
\end{equation}
and the electric field
\begin{equation}
\label{deutschElectricField}
\begin{aligned}
E_r&=-k\frac{mR^2}{r^4}\,\left[\Bigl(\frac{3}{2}\cos2\theta+\frac{1}{2}\Bigr)\cos\theta_m+\frac{3}{2}\sin2\theta\sin\theta_m\cos(\varphi-\varphi_m)\right],\\
E_\theta&=-k\frac{mR^2}{r^4}\,\left[\sin2\theta\cos\theta_m+\Bigl(\frac{r^2}{R^2}-\cos2\theta\Bigr)\sin\theta_m\cos(\varphi-\varphi_m)\right],\\
E_\varphi&=k\frac{mR^2}{r^4}\,\Bigl(\frac{r^2}{R^2}-1\Bigr)\cos\theta\sin\theta_m\sin(\varphi-\varphi_m).
\end{aligned}
\end{equation}
Here, $\theta_m$ and $\varphi_m$ are the polar and azimuthal angles of the magnetic axis specified by the direction of the magnetic dipole moment vector~$\mathbf{m}$, $k=\Omega/c$~is the wave number corresponding to the angular frequency of neutron star rotation~$\Omega$, and $R$~is the stellar radius. The azimuthal angle~$\varphi_m=\Omega t$ was chosen in such a way that its value was zero at~$t=0$. Equations~\eqref{deutschMagneticField} and~\eqref{deutschElectricField} are valid as long as we consider the electromagnetic field at distances much smaller than the light cylinder radius~$R_c=c/\Omega$:
\begin{equation}
\label{quasistaticCondition}
(\Omega r_\perp/c)^2\ll1,
\end{equation}
where $r_\perp$ is the distance from the rotation axis of the neutron star to the point under consideration. We are primarily interested in the regions of the inner vacuum magnetosphere where an efficient one-photon production of electron–positron pairs is possible. For pulsars with a characteristic surface magnetic field of $10^{12}$~G, the characteristic distance at which the pair production is still possible is $\sim(10{-}20)R$; therefore, for typical values of the stellar radius, $R\simeq10$~km, and spin period, $P\sim0.1{-}1$~s, parameter~\eqref{quasistaticCondition} is $\sim10^{-5}{-}10^{-3}$. Under these conditions, Eqs.~\eqref{deutschMagneticField} and~\eqref{deutschElectricField} yield a fairly accurate result.

Using Eqs. \eqref{deutschMagneticField} and \eqref{deutschElectricField} for the electromagnetic field by Deutsch, we can easily derive an equation of the force-free surface~$\mathbf{E}\cdot\mathbf{B}=0$---the surface at each point of which the longitudinal electric field is zero:
\begin{equation}
\label{FFSequation}
r_{ffs}^2=R^2\left(1-4\frac{\cos\theta\cos^2\theta'}{\sin\theta_m\cos\theta''}\right).
\end{equation}
Here, we introduced the angles $\theta'$ and $\theta''$ as follows:
\begin{equation}
\label{thetaAngles}
\begin{aligned}
\cos\theta'&=\cos\theta\cos\theta_m+\sin\theta\sin\theta_m\cos(\varphi-\varphi_m),\\
\cos\theta''&=-\cos\theta\sin\theta_m+\sin\theta\cos\theta_m\cos(\varphi-\varphi_m).
\end{aligned}
\end{equation}
The structure of the force-free surface \eqref{FFSequation} can be understood from Fig.~1. It shows the sections of the force-free surface by the plane~$\varphi-\varphi_m=\{0,\pi\}$ that passes through the rotation and magnetic axes and by the plane~$\varphi-\varphi_m=\{\pi/2,3\pi/2\}$ that also passes through the rotation axis but orthogonally to the preceding plane for various inclinations of the magnetic axis to the rotation axis of the neutron star. An idea of the 3D~structure of the force-free surface can be gotten from Fig.~3 (see below), which shows the particle trajectories lying on this surface.

For an aligned rotator, the force-free surface is just the equatorial plane~$\theta=\theta'=\pi/2$. For an arbitrary oblique rotator, the force-free surface can be divided into two regions. One region consists of two vaulted parts adjacent to the neutron star surface at the points of the equator~$\theta=\pi/2$ and the magnetic equator~$\theta'=\pi/2$. The force-free surface touches the neutron star surface at all points of the magnetic equator, which is not observed at the points of the ordinary equator. The second region is the surface that starts at the magnetic equator and goes to infinity. For an easier association of this surface with the sections by the plane~$\varphi-\varphi_m=\{0,\pi\}$, it is convenient to imagine it as being composed
of two open sheets. One edge of each sheet is adjacent to the magnetic equator and to the straight line passing through the center of the neutron star orthogonally to its rotation and magnetic axes. The other edge of the sheet goes to infinity in such a way that the angle~$\theta''\rightarrow\pi/2$ as~$r\rightarrow\infty$ for an arbitrary point lying on
the force-free surface. At large~$r$, the sheet differs little from the plane~$\theta''=\pi/2$. However, at relatively small distances from the stellar plane, the difference is noticeable and a dome-shaped part ending at the magnetic equator is formed above the plane $\theta''=\pi/2$. The two sheets are smoothly jointed at the points of the straight line~$\theta=\theta'=\theta''=\pi/2$ mentioned above to
form a single sheet.

Below, we will everywhere consider only the values of~$\theta_m$ lying within the range from~$0$ to~$\pi/2$. This does not limit the generality, because a rotator with an angle~$\chi>\pi/2$ between the vectors~$\mathbf{m}$ and~$\mathbf{\Omega}$ is equivalent to a rotator with~$\theta_m=\pi-\chi<\pi/2$; $-\Omega$~should be substituted for~$\Omega$ in all formulas.

\section{The motion of charged particles}

To investigate the motion of particles in a vacuum magnetosphere, we will use the classical Dirac–Lorentz equation
\begin{equation}
\label{DiracLorentzEquation}
m_e\ddot{x}^i=\frac{2}{3c^3}e^2\left[\dddot{x}^i+\frac{1}{c^2}\dot{x}^i\,\ddot{x}^k\ddot{x}_k\right]+F^i,
\end{equation}
where $x^i=(ct,\mathbf{r}^T)^T$ is a contravariant 4-vector containing the time $t$ and coordinates $\mathbf{r}=(x,y,z)^T$ of the particle in the laboratory frame of reference ($T$ is the transposition symbol), $m_e$ is the particle mass, $e$ is the particle charge, and $c$ is the speed of light. The dot over the 4-vector denotes differentiation with respect to the proper time $\tau$ of the particle, i.e., the time in the
comoving frame of reference. The differentials of the time $dt$ in the laboratory frame of reference and the proper time $d\tau$ of the particle are related by the relation $d\tau=dt/\gamma$, where $\gamma$ is the particle Lorentz factor. The 4-force $F^i$ acting to the particle is defined as $F^i=e F^{ik}\dot{x}_k/c$, where
\begin{equation*}
F_{ik}=\partial A_k/\partial x^i-\partial A_i/\partial x^k
\end{equation*}
is the electromagnetic field tensor and the definition of the 4-potential
\begin{equation*}
A^i=(A^0,A^1,A^2,A^3)^T=(\phi,\mathbf{A}^T)^T
\end{equation*}
includes the standard scalar, $\phi$, and vector, $\mathbf{A}$, electromagnetic potentials.

Let us pass to dimensionless variables. We will measure the strength of the electric and magnetic fields in units of the so-called critical field $B_c=m_e^2c^3/e\hbar\approx4.4\times10^{13}$~G,
the particle velocity in units of the speed of light~$c$, the particle charge in units of the positron charge~$e$, the particle mass in units of the electron mass~$m_e$, the particle energy in units of the electron rest mass~$m_e c^2$, all distances in units of the electron Compton wavelength~$^-\!\!\!\!\lambda=\hbar/m_e c\approx3.86\times10^{-11}$~cm, and all times in units of~$^-\!\!\!\!\lambda/c$. We will immediately note that in these units, $1000\text{ km}\approx2.6\times10^{18}$ and~$1\text{ s}\approx7.8\times10^{20}$.

After the reduction to dimensionless form described above and the separation of the scalar and vector components of the 4-vector~$x^i$, the Dirac–Lorentz equation~\eqref{DiracLorentzEquation} is reduced to the system of equations
\begin{alignat}{2}
\label{dimentionlessDLEquation1}
\frac{d\gamma}{dt}&=\frac{2}{3}\alpha\gamma\left[\frac{d^2\gamma}{dt^2}
-\gamma^3\Bigl(\frac{d\mathbf{v}}{dt}\Bigr)^2\right]&&\pm\mathbf{v}\cdot\mathbf{E},
\\
\label{dimentionlessDLEquation2}
\gamma\frac{d\mathbf{v}}{dt}&=\frac{2}{3}\alpha\gamma\left[3\frac{d\gamma}{dt}
\frac{d\mathbf{v}}{dt}+\gamma\frac{d^2\mathbf{v}}{dt^2}\right]
&&\pm\biggl[\mathbf{E}-\mathbf{v}(\mathbf{v}\cdot\mathbf{E})+\mathbf{v}\times\mathbf{B}\biggr],
\end{alignat}
where $\mathbf{v}$ is the particle velocity and $\alpha=e^2/\hbar c\approx1/137$ is the fine-structure constant. When the motions of positrons and electrons are investigated, the plus and minus signs, respectively, should be taken in the system of equations \eqref{dimentionlessDLEquation1} and~\eqref{dimentionlessDLEquation2}. For convenience, we will write the plus sign everywhere below, separately stipulating in important cases what will change if not positrons but electrons are considered.

A charged particle produced in the magnetosphere, be it an electron or a positron, will be subjected to an electric field and will accelerate. It is easy to estimate the time in which the particle motion will become relativistic, $\tau_{rel}\approx 1/E_\parallel$.
From the general form of Eqs.~\eqref{deutschMagneticField} and~\eqref{deutschElectricField}, we see how the electric field strength~$\mathbf{E}$ is
related to the magnetic field strength~$\mathbf{B}$, with the relation~$E_{surf}\approx RB_{surf}/R_c$ holding for the fields on the stellar surface. Since the surface electric field for a typical surface magnetic field $B_{surf}\sim0.01{-}0.10$ and a ratio of
the neutron star radius to the light cylinder radius $R/R_c\sim10^{-4}{-}10^{-3}$ is $E_{surf}\sim10^{-6}{-}10^{-4}$, the particle transition time to the relativistic regime is $\tau_{rel}\sim10^{4}{-}10^{6}$. This means that the particle reaches a relativistic
velocity in a time of $10^{-17}-10^{-15}$~s and subsequently its motion may be considered as ultrarelativistic. Obviously, in this case, the particle will travel a distance of
no more than $10^{4}{-}10^{6}$ Compton wavelengths.

Subsequently, the particle will continue to accelerate already in the ultrarelativistic regime. As is well known, in general, the particle will move along curved magnetic field lines. Acquiring an increasingly high energy, the particle will begin to emit the so-called curvature photons. Intense curvature radiation leads to the loss of particle energy. The emerging force of radiative friction is taken into account by the Dirac–Lorentz equation.

The intensity of curvature radiation increases with Lorentz factor~$\gamma$. Therefore, in the long run, having reached some maximum Lorentz factor~$\gamma_0$, the particle
subsequently will not undergo any acceleration, since the entire energy gained by it through the action of the electric field will be lost through the emission of curvature photons. To determine~$\gamma_0$, we should find a stationary solution of Eq.~\eqref{dimentionlessDLEquation1} by setting $d\gamma/dt=d^2\gamma/dt^2=0$:
\begin{equation}
\label{stationaryECE}
\frac{2}{3}\alpha\gamma_0^4\Bigl(\frac{d\mathbf{v}}{dt}\Bigr)^2=\mathbf{v}\cdot\mathbf{E}.
\end{equation}
Equation \eqref{stationaryECE} shows that in a stationary state, the entire work of the electric field on the particle completely transforms into the energy of curvature radiation. In the case under consideration, the particle velocity vector has a unit length, i.e., $v=1$. Therefore, $d\mathbf{v}/{dt}=\mathbf{n}/\rho$,
where $\mathbf{n}$ is the vector of the principal normal to the particle trajectory and $\rho$ is the radius of curvature of the trajectory. The maximum particle Lorentz factor takes the form
\begin{equation}
\label{gammaMax}
\gamma_0=\left(\frac{3}{2\alpha}E_\parallel\rho^2\right)^{1/4}.
\end{equation}
For a characteristic longitudinal electric field $E_\parallel\sim10^{-4}$ and a radius of curvature of the particle trajectory $\rho\sim R\approx2.6\times10^{16}$, the maximum particle Lorentz factor is $\gamma_0\approx6\times10^7$.

Let us estimate the time it takes for the particle to reach the maximum Lorentz factor $\gamma_0$---the time of its full acceleration and its transition to the quasi-stationary regime of motion defined by the balance condition~\eqref{stationaryECE}. The sought-for time is $\tau_{st}\approx\gamma_0/E_\parallel$. Thus, the particle reaches its maximum energy in a time $\tau_{st}\sim10^{12}$ ($10^{-9}$~s in dimensional units) and subsequently moves in such a way that the work of the electric field on the particle per unit time is equal to the total intensity of curvature radiation. In this case, the particle travels a distance of $\sim10^{12}$ (of the orders of several tens of centimeters in dimensional units), which is much smaller than~$R$. Therefore, the assumption that the electric field does not change in the time $\tau_{st}$ of full particle acceleration is valid. Hence, the electrons and positrons may be assumed to accelerate almost instantaneously at the point where the electron–positron pair was produced.

However, since both the electric field strength and the radius of curvature of the trajectory change as the particle moves, the Lorentz factor $\gamma_0$ also changes with
time. This means that the particle will adjust itself, acquiring energy due to the work of the electric field on it if $\gamma_0$ increases along the trajectory and losing
energy through the emission of curvature photons if $\gamma_0$ decreases along the particle trajectory. Now, it is important to determine the rate of this adjustment of
the particle energy. If this rate exceeds considerably the rate of change in $\gamma_0$ along the trajectory, then the particle Lorentz factor may be assumed to be determined by the coordinates of the point at which the particle is located, because the value of $\gamma_0$ itself at some fixed instant of time depends only on the coordinates of the point under consideration but under no circumstances on the particle velocity.

Let us find the law of change in the particle Lorentz factor $\gamma$ as its stationary value of~$\gamma_0$ is approached. For this purpose, we should use Eq.~\eqref{dimentionlessDLEquation1} by representing the particle Lorentz factor as the sum of its stationary value, $\gamma_0$, and some deviation from it, $\delta\gamma$. The deviation $\delta\gamma$ should be considered small, i.e., $\delta\gamma\ll\gamma_0$. Linearizing Eq.~\eqref{dimentionlessDLEquation1} and eliminating the self-accelerating
solution, we obtain $\delta\gamma=\delta\gamma_i\exp(-t/\tau_0)$.
Thus, when the particle Lorentz factor deviates from its stationary value, $\gamma_0$, by~$\delta\gamma_i$, it will approach $\gamma_0$ exponentially with a decay time constant $\tau_0=3\rho^2/8\alpha\gamma_0^3$. For a characteristic radius of curvature of the trajectory $\rho\sim10^{17}$ and Lorentz factor $\gamma_0\sim10^7{-}10^8$, the decay time is $\tau_0\sim10^{11}{-}10^{14}$ ($10^{-10}{-}10^{-7}$~s in dimensional units). In time $\tau_0$, the particle will travel a distance $l_0=\tau_0$ of $\sim10^{11}{-}10^{14}$ Compton wavelengths (i.e., from centimeters to tens of meters). Now, recall that the characteristic distances at which the electric field and the radius of curvature of the magnetic field lines and, hence, the stationary particle Lorentz factor $\gamma_0$
change are of the order of $R$. Since $l_0\ll R$, the particles will adjust themselves to a change in $\gamma_0$ as they move in the magnetosphere. Thus, the Lorentz factor of the particle may be assumed to be completely determined by its coordinates.

So far, we have discussed the particle energetics using only Eq.~\eqref{dimentionlessDLEquation1}. Let us now investigate the second equation~\eqref{dimentionlessDLEquation2} in more detail and examine how this equation defines the particle trajectory. Let us estimate the terms that arise when the self-action of charged particles is taken into account by assuming that the time $\tau_{st}$ has elapsed after the generation and the particle has already fully accelerated and passed to the quasi-stationary regime of motion defined by condition~\eqref{stationaryECE}.The first term in Eq.~\eqref{dimentionlessDLEquation2} containing the square brackets is equal in order of magnitude to $\alpha\gamma_0^2/R^2$. It is small compared to the second term, because the condition $\alpha\gamma_0^2/R^2\ll E$ is met. This can be verified by taking typical values of the particle Lorentz factor, $\gamma_0\sim10^8$, the stellar radius, $R\sim10^{17}$, and the electric field, $E\sim10^{-4}$. However, the satisfaction of this condition is not a sufficient reason to discard the self-action terms. Indeed, the term on the left-hand side of Eq.~\eqref{dimentionlessDLEquation2} is equal in order of magnitude to $\gamma_0/R\sim10^{-9}$ and is also much smaller than~$E$. For a justified discarding of the terms that allow for the effect of the field generated by charged particles on their proper motion, it is necessary that these terms be small compared to $\gamma_0/R$, $\alpha\gamma_0/R\ll1$. We see that this condition is always met; therefore, the motion of particles in the vacuum magnetosphere of a neutron star will be described by the equation
\begin{equation}
\label{EqOfMotion}
\gamma\frac{d\mathbf{v}}{dt}=
\mathbf{E}-\mathbf{v}(\mathbf{v}\cdot\mathbf{E})+\mathbf{v}\times\mathbf{B}.
\end{equation}
Thus, Eq.~\eqref{EqOfMotion} describes the proper motion of the particle, while its energetics is defined by Eq.~\eqref{dimentionlessDLEquation1}, which is reduced to equality \eqref{gammaMax} in the quasi-stationary case..

The total particle velocity $\mathbf{v}$ will be rewritten as $\mathbf{v}=\mathbf{b}+\mathbf{v}_{e}+\mathbf{v}_{c}$, where $\mathbf{b}=\mathbf{B}/B$ is a unit tangent vector to the magnetic field line, $\mathbf{v}_e=\mathbf{E}\times\mathbf{B}/B^2$ is the electric drift velocity, and $\mathbf{v}_{c}=\gamma\mathbf{b}\times d\mathbf{b}/dt/B$ is the centrifugal drift velocity.

Let us estimate the drift velocities $\mathbf{v}_e$ and~$\mathbf{v}_c$. in order
of magnitude. The electric drift velocity is equal to the ratio of the electric and magnetic field strengths, $v_e\sim E/B$, and is $R/R_c\sim10^{-4}$ in order of magnitude. For $\gamma\sim10^8$, $\rho\sim10^{17}$, and $B\sim0.01{-}0.10$, the centrifugal drift
velocity is $v_c\sim10^{-8}{-}10^{-7}$. Thus, the centrifugal drift velocity is a quantity of the next order of smallness compared to the electric drift velocity; therefore, we will disregard the centrifugal drift. So, we will assume that $\mathbf{v}=\mathbf{b}+\mathbf{v}_e$ to terms $o(kr)$---quantities of the next order of smallness (quadratic or higher order) in $kr\sim R/R_c$.

The electromagnetic field described by Eqs.~\eqref{deutschMagneticField} and~\eqref{deutschElectricField} depends periodically on time. Let $t$, $r$, $\theta$, and~$\varphi$ be the spherical coordinates in the laboratory frame of reference. Let us pass to new coordinates $t'$, $r'$, $\vartheta'$, and~$\varphi'$ using the formulas $t'=t$, $r'=r$, $\vartheta'=\theta$, and~$\varphi'=\varphi-\Omega t$, where $t'$, $r'$, $\vartheta'$, $\varphi'$ are the time and the coordinates in a rotating frame of reference. In these coordinates, the electromagnetic field depends only on $r'$, $\vartheta'$, and $\varphi'$ but not on $t'$. Below, we will omit the primes on the variables in the cases where it is clear that the analysis is performed in the rotating frame of reference. The velocity and acceleration transformations when passing to the rotating frame of reference have a standard form:
\begin{equation}
\label{velocityAccelerationTransform}
\begin{aligned}
\mathbf{v}&=\mathbf{v}'+\mathbf{v}_{tr},\\
\frac{d\mathbf{v}}{dt}&=\frac{d\mathbf{v}'}{dt'}
+2\,\mathbf{\Omega}\times\mathbf{v}'+\mathbf{\Omega}\times\mathbf{v}_{tr},
\end{aligned}
\end{equation}
where $\mathbf{v}'$ is the relative velocity and $\mathbf{v}_{tr}=\mathbf{\Omega}\times\mathbf{r}$ is the translational velocity. Formulas~\eqref{velocityAccelerationTransform} are valid if $\dot{\Omega}\ll\Omega^2$.
This condition is equivalent to the requirement that $\dot{P}\ll1$, which is always met, since $\dot{P}\sim10^{-15}$ in order of magnitude. Substituting Eqs.~\eqref{velocityAccelerationTransform} into Eq.~\eqref{EqOfMotion} yields the following equation of particle motion in the rotating frame of reference:
\begin{equation}
\label{EqOfMotionInRotatingFrame}
\gamma\frac{d\mathbf{v}'}{dt'}=
\mathbf{E}+\mathbf{v}_{tr}\times\mathbf{B}
-\mathbf{v}'(\mathbf{v}'\cdot\mathbf{E})+\mathbf{v}'\times\mathbf{B}.
\end{equation}
Here, $\mathbf{E}$ and $\mathbf{B}$ are the electric and magnetic fields defined by Eqs.~\eqref{deutschElectricField} and~\eqref{deutschMagneticField}, respectively, but now, after the change of variables described above, they depend on the coordinates $r'$, $\vartheta'$, and $\varphi'$. In deriving Eq.~\eqref{EqOfMotionInRotatingFrame}, we neglected the Coriolis, $2\,\mathbf{\Omega}\times\mathbf{v}'$, and translational, $\mathbf{\Omega}\times\mathbf{v}_{tr}$, accelerations compared to the relative acceleration~$d\mathbf{v}'/dt'$. This is possible, because the relative acceleration is mainly the axipetal acceleration that arises from the motion of particles along curved magnetic field lines. Therefore, $|d\mathbf{v}'/dt'|\sim1/R$ and is $\sim10^{-17}$. The Coriolis acceleration is $|2\,\mathbf{\Omega}\times\mathbf{v}'|\sim1/R_c$ in
order of magnitude, which is $\sim10^{-21}$. The translational acceleration is even lower, because the translational velocity $v_{tr}$ is definitely lower than~$v'\sim1$ in the magnetospheric regions under consideration, $r\lesssim10R\ll R_c$, where an efficient one-photon production of pairs is possible. The term $\gamma\,\mathbf{\Omega}\times(2\mathbf{v}'+\mathbf{v}_{tr})$ being discarded, which would be present on the left-hand side of Eq.~\eqref{EqOfMotionInRotatingFrame}, is also small compared to the electric field strength, $E\sim10^{-4}$, since it is $\gamma/R_c\sim10^{-13}$ for~$\gamma\sim10^8$. The Lorentz factor $\gamma$ in Eq.~\eqref{EqOfMotionInRotatingFrame} refers to the laboratory frame of reference.

Below, we will investigate the motion of particles in the rotating frame of reference. We are interested in the question of whether the regions in which the accumulation of a primary plasma is possible exist in the magnetosphere. Let us find the equilibrium positions---the points in moving to any of which a charged particle will remain at it for an unlimited time. The coordinates of the equilibrium positions are determined by the conditions for the particle velocity and acceleration being zero in the rotating frame of reference: $\mathbf{v}'=d\mathbf{v}'/dt'=0$. Substituting this condition into Eq.~\eqref{EqOfMotionInRotatingFrame} yields the equation $\mathbf{E}^{eff}=0$, where the effective electric field is $\mathbf{E}^{eff}=\mathbf{E}+\mathbf{v}_{tr}\times\mathbf{B}$. Its components, which can be easily obtained from Eqs.~\eqref{deutschMagneticField} and~\eqref{deutschElectricField} using the expression for~$\mathbf{v}_{tr}$, in spherical coordinates are
\begin{equation}
\label{effectiveElectricField}
\begin{aligned}
E^{eff}_r&=-kr\left[\frac{R^2}{r^2}B_r\cos\theta+\Bigl(1-\frac{R^2}{r^2}\Bigr)
B_\theta\sin\theta\right],\\
E^{eff}_\theta&=kr\Bigl(1-\frac{R^2}{r^2}\Bigr)\left[\frac{1}{2}B_r\sin\theta+B_\theta\cos\theta \right],\\
E^{eff}_\varphi&=kr\Bigl(1-\frac{R^2}{r^2}\Bigr)B_\varphi\cos\theta,
\end{aligned}
\end{equation}
where $B_r$, $B_\theta$, and $B_\varphi$ are defined by Eq.~\eqref{deutschMagneticField}. The condition $\mathbf{E}^{eff}=0$ gives the following set of equilibrium points:
\begin{equation}
\label{nonisolatedEquiPoints}
\begin{alignedat}{2}
r&=R,&\quad\theta&=\frac{\pi}{2}\qquad\text{(equator)},\\
r&=R,&\quad\theta'&=\frac{\pi}{2}\qquad\text{(magnetic equator)},
\end{alignedat}
\end{equation}
\begin{equation}
\label{isolatedEquiPointsAtLeafs}
\begin{aligned}
\frac{r_{+}^2}{R^2}&=\frac{3+\cos\theta_m}{1-\cos\theta_m}
\qquad\text{(open sheets)},
\\
(\theta,\varphi)&=\left\{\Bigl(\frac{\theta_m}{2},\varphi_m\Bigr),
\Bigl(\pi-\frac{\theta_m}{2},\pi+\varphi_m\Bigr)\right\},
\end{aligned}
\end{equation}
\begin{equation}
\label{isolatedEquiPointsAtDomes}
\begin{aligned}
\frac{r_{-}^2}{R^2}&=\frac{3-\cos\theta_m}{1+\cos\theta_m}
\qquad\text{(vaults)},
\\
(\theta,\varphi)&=\left\{\Bigl(\frac{\pi}{2}+\frac{\theta_m}{2},\varphi_m\Bigr),
\Bigl(\frac{\pi}{2}-\frac{\theta_m}{2},\pi+\varphi_m\Bigr)\right\}.
\end{aligned}
\end{equation}
We see that all equilibrium points are on the force-free surface~\eqref{FFSequation} and are separated into two groups: nonisolated and isolated. All points at the equator and the magnetic equator are nonisolated equilibrium points (see~\eqref{nonisolatedEquiPoints}). Four more isolated equilibrium points are added to these two circumferences: two points with the coordinates given by Eqs.~\eqref{isolatedEquiPointsAtLeafs} are located on the open sheets of the force-free surface and the other two points defined by Eqs.~\eqref{isolatedEquiPointsAtDomes} are located in its vaulted parts whose boundary is adjacent to the equator and the magnetic equator. All four points lie in the plane passing through the rotation and magnetic axes.

The question arises as to whether the equilibrium positions found are stable. In addition, the behavior of the particle trajectories near the force-free surface in
general and near the equilibrium positions in particular is not yet clear. We will give an answer to this question after a detailed study of the particle capture by the force-free surface.

\section{Oscillations of charged particles near the force-free surface}

We will quantitatively investigate the motion of charged particles near the force-free surface in the rotating frame of reference using Eq.~\eqref{EqOfMotionInRotatingFrame}.
Let us choose some point $\mathbf{r}_0$ on the force-free surface and decompose the electric field $\mathbf{E}$, the magnetic field $\mathbf{B}$, and the effective electric field $\mathbf{E}^{eff}$ introduced above near this point as
\begin{equation}
\label{fieldExpansion}
\begin{aligned}
\mathbf{E}&=\mathbf{E}_0+(\mathbf{x'}\cdot\nabla)\mathbf{E}_0,\\
\mathbf{B}&=\mathbf{B}_0+(\mathbf{x'}\cdot\nabla)\mathbf{B}_0,\\
\mathbf{E}^{eff}&=\mathbf{E}^{eff}_0+(\mathbf{x'}\cdot\nabla)\mathbf{E}^{eff}_0,
\end{aligned}
\end{equation}
where $\mathbf{E}_0$, $\mathbf{B}_0$, and $\mathbf{E}^{eff}_0$ are the field strengths at point $\mathbf{r}_0$; and $\mathbf{x'}=\mathbf{r}-\mathbf{r}_0$ is the distance from point $\mathbf{r}$ at which we are interested in the field strengths to point $\mathbf{r}_0$. Decompositions~\eqref{fieldExpansion} are valid, because we consider
the particle motion in a small neighborhood of point $\mathbf{r}_0$---at distances much smaller than the characteristic distances of change in the fields, so that $x'\ll R$.
Before seeking for the complete solution of Eq.~\eqref{EqOfMotionInRotatingFrame}, let us first find its particular solution. We will require that this particular solution describes the particle motion with a constant velocity, $\mathbf{v}'_0=\mathrm{const}$. For this solution to exist, two conditions must be met simultaneously:
\begin{equation}
\label{conditionsForPartSolution}
\begin{aligned}
\mathbf{E}^{eff}_0+\mathbf{v}'_0\times\mathbf{B}_0=0,\\
(\mathbf{v}'_0\cdot\nabla)\mathbf{E}^{eff}_0
+\mathbf{v}'_0\times(\mathbf{v}'_0\cdot\nabla)\mathbf{B}_0=0.
\end{aligned}
\end{equation}

The first equation of system \eqref{conditionsForPartSolution} uniquely defines the velocity component $\mathbf{v}'_\perp$ orthogonal to the magnetic field direction, $\mathbf{v}'_\perp=\mathbf{E}_0^{eff}\times\mathbf{B}_0/B_0^2$, with the longitudinal component $v'_\parallel$ still being arbitrary: $\mathbf{v}'_0=v'_\parallel\mathbf{b}+\mathbf{v}'_\perp$. The components of the velocity $\mathbf{v}'_\perp$ are
\begin{equation}
\label{driftInRotatingFrame}
\begin{aligned}
v'_{\perp r}&=
\Omega r\sin\theta\,b_r b_\varphi\frac{1}{2}\,\Bigl(1-\frac{R^2}{r^2}\Bigr),\\
v'_{\perp \theta}&=
\Omega r\!\left[\cos\theta\,b_r+\Bigl(1-\frac{R^2}{r^2}\Bigr)\sin\theta\,
b_\theta\right]b_\varphi,\\
v'_{\perp \varphi}&=
-\Omega r\!\left[\Bigl(1-\frac{R^2}{r^2}\Bigr)\sin\theta\,\Bigl(\frac{1}{2}b_r^2+b_\theta^2\Bigr)
+\cos\theta\,b_r b_\theta\right],
\end{aligned}
\end{equation}
where $b_r$, $b_\theta$, and $b_\varphi$ are the components of the unit vector $\mathbf{b}$.

The longitudinal component $v'_\parallel$ of the velocity $\mathbf{v}'_0$ is uniquely defined by the second equation of system~\eqref{conditionsForPartSolution}. It is easy to obtain the relation
\begin{equation}
\label{ortogonalityCondition}
\mathbf{v}'_0\cdot\nabla(\mathbf{E}_0\cdot\mathbf{B}_0)=0.
\end{equation}

Since the gradient $\nabla(\mathbf{E}_0\cdot\mathbf{B}_0)$ is directed along the
normal to the force-free surface $\mathbf{E}\cdot\mathbf{B}=0$, the velocity $\mathbf{v}'_0$ lies in the tangent plane drawn through point $\mathbf{r}_0$ of
the force-free surface.

Thus, the particular solution that describes the particle motion with a constant velocity exists. The transverse velocity is given by Eqs.~\eqref{driftInRotatingFrame}, while the longitudinal velocity can be obtained from condition~\eqref{ortogonalityCondition} and is
\begin{equation*}
\label{longitudinalVelocity}
v'_\parallel=-\frac{\mathbf{v}'_\perp\cdot\nabla(\mathbf{E}_0\cdot\mathbf{B}_0)}
{\mathbf{b}\cdot\nabla(\mathbf{E}_0\cdot\mathbf{B}_0)}.
\end{equation*}

We see that the particle velocity $\mathbf{v}'_0$ lies in the tangent plane. Therefore, the particle cannot leave the force-free surface and its trajectory lies entirely on this surface. The velocity will change from point to point and is defined by the same expression $\mathbf{v}'_0=v'_\parallel\mathbf{b}+\mathbf{v}'_\perp$, but the field strengths at the point where the particle is located at a given instant of time should be taken every time as $\mathbf{E}_0^{eff}$ and~$\mathbf{B}_0$.

The complete solution of Eq.~\eqref{EqOfMotionInRotatingFrame} can be represented as
\begin{equation}
\label{solutionForm}
\mathbf{x}'=\mathbf{x}'_0+\mathbf{x}'_1,\qquad
\mathbf{v}'=\mathbf{v}'_0+\mathbf{v}'_1,
\end{equation}
where $\mathbf{x}'_0$ and $\mathbf{v}'_0$ are the adiabatic solution found
above. No special constraints, except the smallness of~$\mathbf{x}'_1$ compared to the characteristic scales of change in the electromagnetic field equal to $R$, are imposed on $\mathbf{x}'_1$ and $\mathbf{v}'_1$ a priori. In particular, the velocity $\mathbf{v}'_1$ can be close to the speed of light. Substituting equalities~\eqref{solutionForm} into~\eqref{EqOfMotionInRotatingFrame} yields the equation
\begin{equation}
\label{EqOfMotionNearFFS}
\gamma\frac{d\mathbf{v}'_1}{dt'}=
(\mathbf{x}'_1\cdot\nabla)\mathbf{E}^{eff}_0+\mathbf{v}'_0\times(\mathbf{x}'_1\cdot\nabla)\mathbf{B}_0
+\mathbf{v}'_1\times\mathbf{B}-\mathbf{v}'_1(\mathbf{v}'_1\cdot\mathbf{E}),
\end{equation}
where $\mathbf{E}$ and $\mathbf{B}$ are defined by Eqs.~\eqref{fieldExpansion}, while the coordinate $\mathbf{x}'$ in them is defined by~\eqref{solutionForm}. In deriving Eq.~\eqref{EqOfMotionNearFFS}, we used the condition $v'_0\ll1$. The term $\mathbf{v}'_1(\mathbf{v}'_1\cdot\mathbf{E})$ is retained, because, in general, the velocity $\mathbf{v}'_1$ can be of the order of unity; therefore, the term under consideration can be equal to $\mathbf{E}$ in order of magnitude.

To derive an equation for describing the oscillations of the particle with no special constraints imposed on its velocity, we must use Eq.~\eqref{EqOfMotionNearFFS} by assuming that $\mathbf{v}'_1\parallel\mathbf{B}$ and write the derived equation in projection onto the magnetic field direction:
\begin{equation}
\label{relativisticOscillationEq1}
\gamma\frac{dv'_1}{dt'}=-\omega^2x'_1-(v'_1)^2E_\parallel.
\end{equation}
Here, $\omega$ is the frequency of nonrelativistic oscillations defined by the formula
\begin{equation}
\label{nonrelativisticOscillationsFrequency}
\omega^2=-\frac{\mathbf{b}\cdot\nabla(\mathbf{E}_0\cdot\mathbf{B}_0)}{\mathbf{B}_0}.
\end{equation}

Let us first consider the nonrelativistic case of particle oscillations. Since the term $(v'_1)^2E_\parallel$ at $v'_1\ll1$ is negligibly small and $\gamma\approx1$, Eq.~\eqref{relativisticOscillationEq1} transforms to an ordinary equation of nonrelativistic oscillations with frequency~$\omega$. Let us find the applicability criterion for the nonrelativistic approximation. As we see from Eq.~\eqref{nonrelativisticOscillationsFrequency}, the characteristic oscillation frequency is $\omega\sim\sqrt{\Omega B}$ or, equivalently, $\omega\sim\sqrt{B/R_c}$.
For a characteristic light cylinder radius $R_c\sim10^{20}$ (for pulsars with
a period $P\sim1$~s) and magnetic fields $B\sim0.01{-}0.10$, the frequency of nonrelativistic oscillations is $\omega\sim10^{-11}{-}10^{-10}$ (corresponding to a frequency $\nu=\omega/2\pi\sim1{-}10\text{ GHz}$ in dimensional units). Over the oscillation period, the particle will traverse a distance that is definitely smaller than $1/\nu$ and the maximum amplitude of nonrelativistic oscillations is $l_{nro}\approx 1/\omega$. The oscillatory motion will be nonrelativistic if $x'_1\ll l_{nro}$. For the
frequencies $\omega$ found above, we have $l_{nro}\sim10^{10}{-}10^{11}$ ($0.1{-}1.0$~cm in dimensional units). We see that the oscillations of a charged particle near the force-free surface with an amplitude exceeding $1$~cm are definitely relativistic. This suggests that considering the relativistic case of oscillations that is actually realized is of greatest interest.

Let now the velocity $v'_1$ be not low. We must then use the condition $x'_1\ll R$ and, according to Eq.~\eqref{nonrelativisticOscillationsFrequency}, write the longitudinal electric field $E_\parallel=-\omega^2x'_1$. Taking into account the equality $\gamma\approx1/\sqrt{1-(v'_1)^2}$, we obtain an equation for the relativistic oscillations of a charged particle near the force-free surface:
\begin{equation}
\label{absRelativisticOscillationEq}
\frac{dv'_1}{dt'}=-\frac{\omega^2}{\gamma^3}x'_1.
\end{equation}
This equation has the first integral
\begin{equation}
\label{firstIntegral}
C=\gamma+\frac{\omega^2(x'_1)^2}{2}.
\end{equation}
This first integral is equal to the maximum Lorentz factor $\gamma_{\max}$ that a charged particle reaches when passing through the force-free surface, i.e., $C=\gamma_{\max}$, when $x'_1=0$. The first integral $C$ can also be expressed in terms of the oscillation amplitude $A$: $C=1+\omega^2A^2/2$.

For the convenience of our subsequent analysis, will make the following changes of variables:
\begin{equation}
\label{changeOfVariables}
\begin{aligned}
\sin{\phi_0}=\frac{\kappa}{a},\qquad\sin{\phi}=\frac{\omega}{\sqrt{2C}}\frac{x'_1}{a},
\\
\kappa=\sqrt{\frac{C-1}{C+1}},\qquad a=\sqrt{\frac{C-1}{C}}.
\end{aligned}
\end{equation}
If we introduce a function
\begin{equation*}
R(\phi,\kappa)=E(\phi,\kappa)-\cos^2\!\phi_0F(\phi,\kappa),
\end{equation*}
then the relation between $\phi$ and $t$ is given by the equation
\begin{equation}
\label{phitRelation}
t=\frac{\sqrt{2C}}{\omega}\frac{R(\phi,\kappa)}{\sin\phi_0}.
\end{equation}
Here, $F(\phi,\kappa)$ and $E(\phi,\kappa)$ are the elliptic integrals of the first and second kinds, respectively.

The oscillation period can be accurately determined at once by noting that $\phi=0$ corresponds to an equilibrium position, while $\phi=\pi/2$ corresponds to a turning point. Since the particle travels the distance from the equilibrium position $x'_1=0$ to the turning point $x'_1=A$ in a quarter of the period,
\begin{equation*}
\label{accurateT}
T=4\frac{\sqrt{2C}}{\omega}\frac{\mathbf{R}(\kappa)}{\sin\phi_0},
\end{equation*}
where $\mathbf{R}(\kappa)=\mathbf{E}(\kappa)-\cos^2\!\phi_0\mathbf{K}(\kappa)$, while
$\mathbf{K}(\kappa)=F(\pi/2,\kappa)$ and $\mathbf{E}(\kappa)=E(\pi/2,\kappa)$ are the complete elliptic integrals of the first and second kinds.

The exact solution of the oscillation equation~\eqref{absRelativisticOscillationEq} can be written as
\begin{equation}
\label{oscSolution}
x'_1=A\sin\mathcal{Q}\biggl(\frac{\omega t}{\sqrt{2(C+1)}}\biggr).
\end{equation}
Here, we introduced a function $\mathcal{Q}(z)$ that is the inverse of the function $R(\phi,\kappa)$, so that $z=R(\mathcal{Q}(z),\kappa)$ for any real number $z\in\mathbb{R}$. This can be done, because the function $R(\phi,\kappa)$ strictly increases and is continuously differentiable with respect to the variable $\phi$ on the
entire real $\mathbb{R}$ axis at $0\leqslant\kappa<1$, with $R(\mathbb{R},\kappa)=\mathbb{R}$. When these conditions are met, the inverse function $\mathcal{Q}(z)$ exists and is also a single-valued, strictly increasing and continuously differentiable function on the $\mathbb{R}$ axis, so that $\mathcal{Q}(\mathbb{R})=\mathbb{R}$. Its derivative $d\mathcal{Q}/dz$ nowhere becomes neither zero nor infinite, since this is true for the partial derivative $\partial R(\phi,\kappa)/\partial\phi$ at all $C$ from the interval $1\leqslant C<\infty$ of interest to us (this corresponds to any physically possible values of the particle Lorentz factor $\gamma_{\max}$).

Equation \eqref{oscSolution} exhausts the problem of charged particle oscillations near the force-free surface, provided that there are no energy losses through radiation. The asymptotic limits of the function $\mathcal{Q}(z)$ at values of $\kappa$ close to zero and unity are
\begin{equation}
\label{qAsymptotics}
\mathcal{Q}(z)=
\begin{cases}
\begin{alignedat}{2}
&2\,z,&\qquad \kappa&=0,\\
&\arcsin (z-2\,\mathrm{h}(z))+\pi\,\mathrm{h}(z),&\qquad \kappa&\rightarrow1,
\end{alignedat}
\end{cases}
\end{equation}
where the function $\mathrm{h}(z)$ taking on integer values is defined by the formula
\begin{equation*}
\label{hz}
\mathrm{h}(z)=\left[\frac{z+1}{2}\right].
\end{equation*}
Here, the integer part of a real number $y$ is denoted by $[y]$. No constraints are imposed on $z$ in Eq.~\eqref{qAsymptotics}. In the nonrelativistic case of oscillations, we must use the asymptotic limit of the function $\mathcal{Q}(z)$
at $\kappa=0$. As we see from Eq.~\eqref{oscSolution}, the equality $z=\omega t/2$ holds at $C\simeq1$. We then immediately obtain the equation of harmonic oscillations $x'_1=A\sin(\omega t)$ with frequency $\omega$. In the ultrarelativistic case, $C\gg1$, which corresponds to $\kappa\rightarrow1$, and substituting the corresponding asymptotic limit of $\mathcal{Q}(z)$ \eqref{qAsymptotics} into Eq.~\eqref{oscSolution} gives
\begin{equation}
\label{relOscillations}
x'_1=A(-1)^{\,\mathrm{h}(z)}\left(z-2\,\mathrm{h}(z)\right),
\end{equation}
where $z=t/A$. This equation describes saw-tooth oscillations with amplitude $A$ and period $T=4A$.

\section{Energy losses during oscillations}

During its oscillations near the force-free surface, a charged particle will lose its energy through radiation. Let us investigate the time dependence of the rate of decrease in particle energy. The particle energy is characterized by the first integral $C$ found above (see \eqref{firstIntegral}). The change in energy in the ultrarelativistic case where $C\gg1$ is of greatest interest to us, since in the nonrelativistic case, $C\simeq1$ and is almost constant.

Let us turn to the Dirac–Lorentz equation \eqref{dimentionlessDLEquation1}. Using the fact that $v'_1\simeq1$, we obtain
\begin{equation}
\label{DLEforUltrarelativisticOscillations}
\frac{dC}{dt}=\frac{2}{3}\alpha\gamma\!\left[\frac{d^2\gamma}{dt^2}
-\frac{\gamma^3}{\rho^2}\right].
\end{equation}
The rate of decrease in $C$ will not be constant on timescales of the order of the period of particle motion, because it depends on the specific particle Lorentz factor $\gamma$, whose value, obviously, changes significantly in a quarter of the oscillation period from $1$ at the turning point $x'_1=A$ to $C\gg1$ at $x'_1=0$. Nevertheless, this will not be an obstacle to solving the problem of change in particle energy if the value of $C$ itself changes little over the oscillation period, $dC/dt\ll C/T$. When this adiabatic approximation holds, Eq.~\eqref{DLEforUltrarelativisticOscillations} can be carefully
averaged over the oscillation period. In this case, we must average the expressions dependent on the particle coordinate $x'_1$ not formally but using the explicit time
dependence \eqref{relOscillations} of the coordinate of the oscillating
particle $x'_1(t)$, which is an exact one at $C\gg1$.

After its averaging, the differential equation for $C$ transforms to
\begin{equation}
\label{averagedDLEforUltrarelativisticOscillations}
\frac{dC}{dt}=-\frac{4}{9}\alpha\omega^2C
\left[1+\frac{64}{105}\,\frac{C^3}{\omega^2\rho^2}\right].
\end{equation}
In deriving Eq.~\eqref{averagedDLEforUltrarelativisticOscillations},we neglected the second derivative of the energy $C$ with respect to time. This action can be justified rigorously only a posteriori, because this requires the knowledge of some facts about the structure of the solution $C(t)$. Since this justification, of little interest compared to the main results, is cumbersome, we do not provide it here. With the above stipulations, the exact solution of Eq.~\eqref{averagedDLEforUltrarelativisticOscillations} is
\begin{equation}
\label{cSolution}
\varepsilon(t)=\left[\left(1+\varepsilon_0^{-3}\right)e^{3t/\tau_{d}}-1\right]^{-1/3}.
\end{equation}
where $\varepsilon(t)=C(t)/C_{curv}$ is the normalized first integral, with $\varepsilon_0=\varepsilon(0)$ being its value at the initial time $t=0$. Here, we also introduced the quantity
\begin{equation}
\label{Ccurv}
C_{curv}=\frac{105^{\,1/3}}{4}(\omega\rho)^{2/3}\approx1.2\:(\omega\rho)^{2/3}
\end{equation}
and the decay time constant $\tau_{d}=9/4\alpha\omega^2$. The physical meaning of the constants $C_{curv}$ and $\tau_d$ can be easily understood by investigating the behavior of solution \eqref{cSolution} at various initial values of $\varepsilon_0$. Let first $\varepsilon_0\ll1$. The solution of Eq.~\eqref{averagedDLEforUltrarelativisticOscillations} is then given by the formula
\begin{equation}
\label{undercurvSolution}
\varepsilon(t)=\varepsilon_0e^{-t/\tau_d}\qquad(C_0\ll C_{curv}).
\end{equation}
We see that the case of $C_0\ll C_{curv}$ corresponds to discarding the term in brackets in Eq.~\eqref{averagedDLEforUltrarelativisticOscillations} proportional to $(C/C_{curv})^3$. However, this means that the energy losses by a charged particle through the emission of curvature photons by it are small compared to the losses due to the presence of a term proportional to $\gamma\,d^2\gamma/dt^2$ in Eq.~\eqref{DLEforUltrarelativisticOscillations}. These energy losses can be associated with the bremsstrahlung losses attributable to a pure oscillatory particle motion. They are nonzero only if the particle energy changes with time. In contrast to them, the curvature losses depend only on the radius of curvature of the particle trajectory and the particle energy. Obviously, the intensity of the curvature losses does not depend on the time derivatives of the charged particle energy. Thus, at $C_0\ll C_{curv}$ the
particle energy decreases exponentially with the decay time constant $\tau_d$, while the losses themselves are attributable mainly to bremsstrahlung.

Let now $\varepsilon_0\gg1$. Obviously, in this case, the behavior of $C$ on relatively short time scales $t$ is of greatest interest, because $C$ decreases with time and becomes smaller than $C_{curv}$ on fairly long time scales, so that the decay becomes exponential. On short time scales $t$, we have
\begin{equation}
\label{tauP}
\begin{split}
\varepsilon(t)=\varepsilon_0\left(1-\frac{t}{\tau_{p}}\right),
\qquad\tau_{p}=\frac{\tau_d}{\varepsilon_0^3},
\\
C_0\gg C_{curv},\qquad t\ll\tau_{p}.
\end{split}
\end{equation}
We see that if $C_0\gg C_{curv}$, the particle energy decreases linearly on time scales much shorter than $\tau_p$. However, when $t$ becomes comparable to $\tau_p$, the decay becomes a power-law one:
\begin{equation*}
\label{curvIntermediateTimeSolution}
\varepsilon(t)=\left(3\frac{t}{\tau_d}\right)^{-1/3},
\qquad C_0\gg C_{curv},\qquad\tau_{p}\ll t\ll\tau_d.
\end{equation*}
Here, the meaning of the time $\tau_p$ is clarified as the time in which the initially linear decrease in charged particle energy becomes a power-law one. When a time of the order of $\tau_d$ is reached, the dependence changes again and becomes exponential:
\begin{equation}
\label{curvLargeTimeSolution}
\varepsilon(t)=e^{-t/\tau_d},
\qquad C_0\gg C_{curv},\qquad t\gtrsim\tau_d.
\end{equation}

In contrast to the previously considered case of \eqref{undercurvSolution}, at $C_0\gg C_{curv}$, the energy losses of the particle are attributable mainly to the emission of curvature photons by it. This process dominates on time scales $t\lesssim\tau_d$ and has a power-law pattern. At $t\gtrsim\tau_d$, the particle energy decay becomes exponential, indicative of the particle transition to a regime where the bremsstrahlung related to the acceleration that the particle undergoes due to its oscillatory motion makes a major contribution to the energy losses. Clearly, the transition itself occurs on time scales of the order of $\tau_d$. However, we will make a certain refinement by finding the time $\tau_{curv}$ at which $C=C_{curv}$, i.e., $\varepsilon(\tau_{curv})=1$, using Eq.~\eqref{cSolution}, $\tau_{curv}=\ln2\tau_d/3\approx 0.23\,\tau_d$.

It should be noted that $\tau_{curv}$ is completely independent of the initial particle energy $\varepsilon_0$. This suggests that a charged particle with an initial energy $\varepsilon_0\gg1$, even if very high, will lose the bulk of it in time $\tau_{curv}$. Subsequently, the particle will have an energy $\varepsilon=1$ at time $t=\tau_{curv}$ and an ordinary exponential decay will then take place with the time constant $\tau_d$.
It is not surprising that the dependence $\varepsilon(t)$ (see \eqref{curvLargeTimeSolution}) on time scales $t\gtrsim\tau_d$ does not contain the initial particle energy either. This is because $\tau_{curv}$ is approximately a factor of $4$ shorter than $\tau_d$, while $\varepsilon$ becomes equal to unity in time $\tau_{curv}$, irrespective of $\varepsilon_0$. On long time scales, the relation $\varepsilon(\tau_{curv})=1$ begins to act as the initial condition for the subsequent exponential decay, as is shown by Eq.~\eqref{curvLargeTimeSolution}.

All of the results that we obtained by averaging Eq.~\eqref{DLEforUltrarelativisticOscillations} are valid when the condition of the adiabatic approximation is met. As follows from Eq.~\eqref{averagedDLEforUltrarelativisticOscillations}, this condition gives an upper limit on the charged particle oscillation amplitude:
\begin{equation}
\label{amplitudeRestriction}
A\ll A_{\max}=\left(\frac{945}{128\alpha}\right)^{1/7}\omega^{-6/7}\rho^{2/7}
\approx2.6\;\omega^{-6/7}\rho^{2/7}.
\end{equation}
The condition of the adiabatic approximation can also be interpreted as follows. The oscillation period must be definitely shorter than the characteristic charged particle energy decay time. As such, we must take the fastest time of change in energy that, in our case, is the transition time $\tau_p$ to a power-law decay. Indeed, if we write the condition $T\ll\tau_p$, then the upper limit on the oscillation amplitude following from it will closely coincide with condition \eqref{amplitudeRestriction}.

Finally, let us consider the energy losses of a charged particle in the case of nonrelativistic oscillations that is realized when the oscillation amplitude eventually becomes smaller than $l_{nro}$. In this case, the time dependence of the coordinate is
\begin{equation*}
x'_1=A_0e^{-t/\tau_{nro}}\cos\omega t,
\end{equation*}
where $A_0$ is the initial amplitude and the decay time constant is $\tau_{nro}=3/\alpha\omega^2$. As would be expected, it is equal in order of magnitude to the decay time constant $\tau_{d}$ for the ultrarelativistic case of oscillations.

Let us now estimate all of the quantities introduced above that characterize the time dependence of the energy of a charged particle oscillating near the force-free surface in order of magnitude. First, let us estimate the energy $C_{curv}$, $C_{curv}\sim\left(BR^2/R_c\right)^{1/3}$. For a characteristic surface magnetic field $B\sim0.01-0.10$, a neutron star radius $R\sim10^{17}$, and a light cylinder radius $R_c\sim10^{19}{-}10^{20}$ (we will use these values everywhere for
our estimations below), $C_{curv}$ is $\sim10^{4}$. We can also introduce the corresponding amplitude
\begin{equation*}
A_{curv}=\sqrt{2C_{curv}}/\omega\sim\left(RR_c/B\right)^{1/3}
\end{equation*}
for the same magnetic field, neutron star radius, and light cylinder radius: $A_{curv}\sim10^{12}{-}10^{13}$ (approximately $\sim1$~m in dimensional units).

The characteristic exponential decay time constant, $\tau_d\sim R_c/\alpha B$, is $10^{22}{-}10^{24}$ ($\sim10{-}10^3$~s in dimensional units). The time $\tau_{curv}$ is also of the same order of magnitude. Note that the time $\tau_d$ exceeds the characteristic pulsar spin periods $P$. Nevertheless, if the fields become comparable to critical fields $B\sim1$, which is true for magnetars, then the time $\tau_d$ can be even shorter than the neutron star spin period. Note, incidentally, that the ratio $\tau_d/P$ depends only on the surface magnetic field $B$ but does not depend on $P$.

The order of magnitude of $\tau_p$ is not fixed so clearly, because it depends on the initial particle energy $\varepsilon_0$. However, the range in which $\tau_p$ lies can be determined easily. The upper boundary is given by the condition $\varepsilon_0\gg1$ or, equivalently, $C_0\gg C_{curv}$, when it makes sense to introduce the concept of time $\tau_p$. This means that, in any case, $\tau_p\ll\tau_d$. We will determine the lower
boundary by taking the maximum possible energy $C_0$. Obviously, this value is definitely lower than the maximum achievable Lorentz factor $\gamma_0$ that a particle far from the force-free surface will have. Using the fact that $\gamma_0\sim10^{8}$ and assuming that $\varepsilon_0\sim\gamma_0/C_{curv}\sim10^{4}$, we will find that the inequality $\tau_p\ggg10^{-12}\tau_d$ necessarily holds. This lower limit was estimated with a margin, since, in reality, an oscillating particle cannot have an energy $C_{curv}\sim10^8$. The reason is that even if the particle has such an energy far from the force-free surface, its energy after its capture will be slightly lower, since the particle loses part of its energy as it approaches the force-free surface.

Let us also find the order of magnitude of the upper limit $A_{\max}$ on the oscillation amplitude $A$ when the adiabatic approximation is still applicable. Using Eq.~\eqref{amplitudeRestriction} for this purpose, we will obtain $A_{\max}\sim\left(R^2R_c^3/B^3\right)^{1/7}$. Thus, $A_{\max}\sim10^{13}{-}10^{14}$ (of the order of several tens of meters in dimensional units). An oscillation period $T_{\max}\sim A$ corresponds to this amplitude, which corresponds to $T_{\max}\sim10^{-8}{-}10^{-7}$~s and a frequency $\nu\sim10{-}100$~MHz in dimensional units. Using the amplitude $A_{\max}$, we easily find the maximum oscillation energy
$C_{\max}\sim\left(BR^4/R_c\right)^{1/7}$, which is $C_{\max}\sim(3\times10^{6}){-}10^7$. This allows us to refine the lower limit for the time $\tau_p$. For this purpose, we should take a ratio $C_{\max}/C_{curv}\sim10^2-10^3$  as the maximum energy $\varepsilon_0$. We will then obtain a more realistic lower limit, $\tau_p\gg10^{-9}\tau_d$.

\section{The capture of charged particles}

Above, we found that a charged particle (an electron or a positron) far from the force-free surface moves in such a way that the Lorentz factor of the particle is determined only by its coordinates and is $\gamma_0$ (see \eqref{gammaMax}). However, near the force--free surface, the electric field is weak and the particle energy adjustment time $\tau_0$ is comparable to the characteristic distance (in dimensionless units) of change in field---in the case under consideration, the distance to the force-free surface. This distance $l_c$ can be determined self-consistently from the equation $\tau_0=l_c$; we should express the longitudinal electric field $E_\parallel$ on which $\tau_0$ depends in terms of the same distance $l_c$ measured from the force-free surface along the magnetic field line along which the particle moves. We will use the linear approximation by assuming that $l_c$ is small enough: $E_\parallel(l)=-\omega^2l$. Here, we formally introduced the quantity $\omega^2=-dE_\parallel/dl$; the derivative $dE_\parallel/dl$ should be taken at the point of intersection of the magnetic field line along which the particle moves with the force-free surface. It is easy to verify that $\omega$ is the frequency of nonrelativistic particle oscillations and is defined by Eq.~\eqref{nonrelativisticOscillationsFrequency}. Subsequently, we immediately obtain the capture length $l_c$:
\begin{equation*}
\label{captureLength}
l_c=\left(\frac{3}{512\alpha}\right)^{1/7}
\omega^{-6/7}\rho^{2/7}\approx\omega^{-6/7}\rho^{2/7}.
\end{equation*}

Note that the capture length coincides, to a coefficient, with the maximum oscillation amplitude $A_{\max}$ \eqref{amplitudeRestriction} defined by the condition of the adiabatic approximation. This gives an insight into how the transition from the ultrarelativistic quasi-stationary motion of a particle far from the force-free surface defined by the energy balance condition \eqref{stationaryECE} to the oscillatory motion
of the particle near the force-free surface occurs. For this purpose, we can use qualitative considerations. The capture length $l_c$ is almost a factor of $3$ smaller
than the amplitude $A_{\max}$. This gives us reason to believe that when captured, the particle passes almost immediately to the regime of adiabatic oscillations. To verify this fact, let us estimate the capture amplitude $A_c$---the maximum distance to which the particle will be deflected from the force-free surface when passing through it. It is implied that the charged particle was initially produced far from the force-free surface, started moving toward it, and crosses this surface for the first time. Consider the particle that has not yet crossed the force-free surface at distance $l_c$ from it. On
the one hand, we can estimate the particle Lorentz factor $\gamma_c$ at a given point from Eq.~\eqref{gammaMax}, using the relation $E_\parallel(l)=-\omega^2l$, and setting $l=l_c$, $\gamma_c=4\,\omega^2l_c^2$. On the other hand, we can calculate the first integral \eqref{firstIntegral} by setting $\gamma=\gamma_c$ and $x'_1=l_c$, $C=9\omega^2l_c^2/2$. Using the relation $C=1+\omega^2A^2/2$, we obtain $A_c=3\,l_c\approx2.9\,\omega^{-6/7}\rho^{2/7}$. We see that the capture amplitude $A_c$ virtually coincides with the maximum amplitude of adiabatic oscillations $A_{\max}$ \eqref{amplitudeRestriction}. Consequently, it can be assumed that, having passed through the force-free surface and having been deflected to the distance $A_c$, the charged particle will be captured and subsequently will begin to execute adiabatic oscillations, which we investigated in detail above. For oscillations with the initial amplitude $A_c$, the time $\tau_p$ reaches its lower limit, $\tau_p\sim(10^{-9}-10^{-7})\tau_d$.

For completeness, let us find the frequency of nonrelativistic oscillations $\omega$. Without providing specific calculations, which are fairly cumbersome, we will write out the final result:
\begin{equation}
\label{nonrelativisticFrequencyAtFFS}
\omega^2=-\frac{km}{R^3}\left(\frac{R}{r_{ffs}}\right)^5 4\cos\theta\cos\theta'\,
\frac{2\cos^2\theta'+3}{3\cos^2\theta'+1},
\end{equation}
where $r_{ffs}$ is defined by Eq.~\eqref{FFSequation} of the force-free surface. We will also give an expression for the radius of curvature of the field lines:
\begin{equation}
\label{curvatureRadius}
\rho=\frac{r\,(1+3\cos^2\theta')^{3/2}}{3\sin\theta'(1+\cos^2\theta')}.
\end{equation}
If $\rho$ is required to be calculated at some point on the force-free surface, then, naturally, we should set $r=r_{ffs}$ in this formula.

In the entire previous discussion, we generally paid absolutely no attention on precisely which particle, an electron or a positron, oscillates near the force-free surface. All of the expressions derived above suggested that the oscillating particle had a positive charge. Let us turn to Eq.~\eqref{nonrelativisticFrequencyAtFFS} for the frequency of nonrelativistic oscillations. In the cases where $\omega^2>0$, positron oscillations with frequency $\omega$ take place. If, alternatively, $\omega^2<0$, then electron oscillations take place. Obviously, the frequency of nonrelativistic electron oscillations in this case is $\sqrt{-\omega^2}$. The sign of the oscillating particle charge may be said to coincide with that of $\omega^2$.

As follows from Eq.~\eqref{nonrelativisticFrequencyAtFFS}, the frequency of nonrelativistic oscillations becomes zero only if $\theta=\pi/2$ or $\theta'=\pi/2$. The former and the latter equalities correspond to the equator and magnetic equator of the neutron star, respectively. The equator and the magnetic equator separate the force-free surface into regions in each of which the oscillations of particles only with the same sign can take place. When passing through the line formed by the nonisolated equilibrium positions, the sign of the particle charge is reversed. The simultaneous fulfillment of the equalities $\theta=\pi/2$ and $\theta'=\pi/2$ defines a straight line at each point of which $\omega=0$ as well. However, the signs of the oscillating charges coincide on the open sheets of the force-free surface that are separated by this straight line. If $\theta_m<\pi/2$, then positrons accumulate in the vaulted parts of the force-free surface, while electrons accumulate on its open sheets. The signs of the particle charges are shown in Fig. 1. Note that $\omega^2$ depends on the wave number $k$ and, hence, on its sign. If $\theta_m>\pi/2$, then it is formally necessary to consider the case where the angle between the axes of an oblique rotator is $\pi-\theta_m$ but, at the same
time, to change the sign of the angular velocity $\Omega$. Consequently, in this case, the signs of the charges will be reversed and electrons and positrons will accumulate in the vaulted parts and on the open sheets, respectively.

\section{The trajectories of particles on the force-free surface}

We saw that the motion of a particle captured by the force-free surface is the sum of the drift motion of some guiding center over the force-free surface and the oscillatory motion of the particle about this guiding center. Let us find the exact trajectories of the guiding center in the rotating frame of reference.

Assuming that $\mathbf{v}'_0=v'_\parallel\mathbf{b}+\mathbf{v}'_\perp$, it is easy to verify that $\mathbf{v}'_0\cdot\mathbf{E}^{eff}=0$. We see that the velocity vector $\mathbf{v}'_0$ of the guiding center is always orthogonal to the effective electric field $\mathbf{E}^{eff}$. Let us find the equation of this surface at each point of which the vector $\mathbf{E}^{eff}$ is directed along the normal to the surface in question. If there existed such a function $\xi$ that $\mathbf{E}^{eff}=\nabla\xi$, then the set of
surfaces to which the vector field $\mathbf{E}^{eff}$ is orthogonal would be specified by the equipotential surfaces of the function $\xi$. Directly integrating the equation $\mathbf{E}^{eff}=\nabla\xi$ and taking into account Eqs.~\eqref{effectiveElectricField}, we can make sure that such a function exists:
\begin{equation}
\label{potentialXi}
\xi=kR^2\frac{m}{r^3}\left[\left(\cos\theta_m-\cos\theta\cos\theta'\right)
\left(\frac{r^2}{R^2}-1\right)+\frac{2}{3}\cos\theta_m\right].
\end{equation}
The potential $\xi$ is defined to an arbitrary real constant that we do not write for short. The equation of the surfaces to which the vector field $\mathbf{E}^{eff}$ is orthogonal is $\xi=\mathfrak{C}$, where $\mathfrak{C}=\mathrm{const}$ is an arbitrary real number.

Finding the trajectories of the guiding center is reduced to finding the intersection of the force-free surface \eqref{FFSequation} and the set of equipotential surfaces $\xi=\mathfrak{C}$. Let us introduce a function
\begin{equation*}
\zeta(\theta,\varphi)=\xi(r_{ffs}(\theta,\varphi),\theta,\varphi)\frac{R}{km},
\end{equation*}
where $r_{ffs}(\theta,\varphi)$ is defined by Eq.~\eqref{FFSequation} of the force-free
surface and the potential $\xi$ is defined by Eq.~\eqref{potentialXi}. The sought-for trajectories are then implicitly specified by the equation $\zeta(\theta,\varphi)=\mathfrak{C}$, where, as above, $\mathfrak{C}$ is an arbitrary real number at which, of course, the equation has solutions.

Figure 2 shows phase portraits of the trajectories in angular coordinates $\theta$ and $\varphi$ on the surface in the rotating frame of reference for oblique and orthogonal
rotators. In what follows, we assume that the azimuthal angle corresponding to the magnetic axis is zero. We see that there exist no extended trajectories starting and ending at points on the magnetic equator. This differs from the result by Jackson \citep{Jackson1978}. He argued that some of the trajectories on the force-free surface
of an uncharged orthogonal rotator ended at the magnetic equator; as a result, a reverse drift motion to the stellar surface that sweeps out the captured particles exists. This could be an obstacle to the formation of a force-free magnetosphere. However, Jackson assumed that the particle velocity at some point of the force-free surface was equal to the projection of the drift velocity onto the tangent plane at this point. Above, we ascertained that this is not the case.

A 3D image of the trajectories of the guiding center on the force-free surface for an oblique rotator is presented in Fig. 3. For convenience, let us introduce the potentials of isolated equilibrium positions
\begin{equation*}
\label{isolatedPointsPotential}
\zeta_\pm=\mp\frac{1}{3}\frac{(1\mp\cos\theta_m)^{3/2}}
{(3\pm\cos\theta_m)^{1/2}}
\end{equation*}
and the surface potential $\zeta_{R}=(2/3)\cos\theta_m$. In the dome-shaped parts of the force-free surface, the trajectories are in the shape of closed loops surrounding the equilibrium position. The potential $\zeta_-$  corresponds to the equilibrium position itself, while the interval of potentials $\zeta_R<\zeta<\zeta_-$ corresponds to the set of loops. The equilibrium positions at the equator and the magnetic equator are characterized by the surface potential $\zeta_R$. Two different regions can be distinguished on each of the two open sheets in the phase portrait. The first region is similar to the region on the vaults and also consists of closed loops surrounding the equilibrium position. The potential $\zeta_+$ corresponds to the equilibrium position, while the interval of potentials $\zeta_+<\zeta<0$ corresponds to the set of loops. However, the loops themselves become increasingly elongated as $\zeta$ approaches
zero, tending to go to infinity. The potential $\zeta=0$ corresponds to the separatrix. The separatrix is an open trajectory that goes to infinity but, at the same time, encloses the equilibrium position. The second region on each of the open sheets is formed by the set of trajectories with a potential $0<\zeta<\zeta_R$. None of these trajectories lies entirely on the same sheet. As we see from Fig. 3, all trajectories pass from one sheet to the other, crossing the straight line $\theta=\theta'=\pi/2$ and being closed
around the neutron star. As $\zeta$ tends to $\zeta_R$, these trajectories come increasingly close to the magnetic equator.

The direction of the trajectories can be easily determined using Eq.~\eqref{driftInRotatingFrame}. In the half-space $\cos\theta>0$, the motion along the trajectories in the vaulted part of the force-free surface and on the open sheet inside the separatrix $\zeta=0$ is clockwise when viewed from outside (it is implied that the line of sight crossing the force-free surface is directed along the radius toward the stellar center). In the half-space $\cos\theta<0$, the motion along the trajectories in the vaulted part of the force-free surface and on the open sheet inside the separatrix is counterclockwise when viewed from outside. On the open sheets outside the separatrices (for the trajectories crossing the straight line $\theta=\theta'=\pi/2$), the motion
is from the half-space $\sin\varphi<0$ to the half-space $\sin\varphi>0$ when $\cos\theta>0$ and from the half-space $\sin\varphi>0$ to the half-space $\sin\varphi<0$ when $\cos\theta<0$. All of the aforesaid refers to the case of $\theta_m<\pi/2$. If the inclination of the rotator $\chi>\pi/2$, then we should consider a rotator with an inclination $\theta_m=\pi-\chi$, direct its polar axis from which all polar angles are measured opposite to the vector $\mathbf{\Omega}$, and change the sign of $\Omega$ in Eq.~\eqref{driftInRotatingFrame}. In Fig. 3, this corresponds to reversing the direction of the vector $\mathbf{\Omega}$. Obviously, the direction of all trajectories will be reversed.

Thus, all trajectories of the guiding center on the force-free surface are closed and lie in a finite region, except the separatrix $\zeta=0$. As a charged particle moves along its trajectory, the electromagnetic field and, hence, the particle oscillation parameters change. However, the guiding center moves with drift velocities of $\sim10^{-4}$ and the particle oscillation frequency exceeds $10$~MHz even in the case of ultrarelativistic oscillations. Therefore, the particle travels a distance of the order of several millimeters in the oscillation period, which is much smaller than $R$. As a result, the oscillation parameters of the particle as it moves over the force-free surface change adiabatically. It is important to note that the sign of $\omega^2$ does not change in the case of motion along the trajectory (see the discussion after Eq.~\eqref{curvatureRadius}). Hence, if the particle fell on the force-free surface, then it can no longer leave it, because, first, the trajectory of the guiding center is closed and, second, no instability of the particle oscillations along the magnetic field develops due to the constancy of the sign of $\omega^2$.

\section{Discussion}

The dynamics of the motion of electrons and positrons in the inner vacuum magnetosphere of a neutron star can be imagined as follows. A charged particle produced far from the force-free surface reaches a relativistic velocity in a time $\tau_{rel}\sim10^{-17}-10^{-15}$~s and passes to a quasi-stationary regime of motion in a time $\tau_{st}\sim10^{-9}$~s, traversing a distance of $\sim10-100$~cm. The
particle Lorentz factor $\gamma_0\sim10^7-10^8$ \eqref{gammaMax} is then completely determined by the condition of balance between the power of the forces of an accelerating electric field with $E_{\parallel}/B_c\sim10^{-6}-10^{-4}$ (here, we restore the dimensions of quantities) and the intensity of curvature radiation. In this case, the particle moves virtually along the magnetic field line, because the electric drift velocity is $v_e/c\sim10^{-4}$ and the centrifugal drift velocity is even lower. During the motion along the trajectory, the radius of curvature $\rho$ and the longitudinal electric field $E_{\parallel}$ change slowly, leading to an adjustment of the particle Lorentz factor $\gamma_0$. The adjustment time is fairly short, $\tau_0\sim10^{-10}-10^{-7}$~s, and the particle travels a distance of $\sim1\text{ cm}-100\text{ m}$ in this time; its upper limit is reached near the force-free surface. Since this distance is much smaller than the stellar radius, the particle Lorentz factor is, in fact, determined by the particle coordinates. As a charged particle approaches the force-free surface, the quasi-stationarity condition is violated and the particle passes through the force-free surface with a Lorentz factor at the crossing point $C_{\max}\sim10^6-10^7$. Once the particle has crossed the force-free surface, it is deflected to a distance $A_c\sim A_{\max}\sim10-100$~m \eqref{amplitudeRestriction}. Subsequently, the particle begins to execute adiabatic ultrarelativistic oscillations with a frequency $\nu\sim10-100$~MHz. The oscillations decay due to the energy losses through radiation and their frequency increases. The oscillation energy decreases initially linearly and, in a time $\tau_p\sim10^{-8}-10^{-4}$~s (see \eqref{tauP}), a power-law decay through curvature losses begins. On time scales of $\tau_{curv}\sim\tau_d\sim10-1000$~s, the decay becomes exponential with the time constant $\tau_d$; bremsstrahlung begins to make a major contribution to the energy losses. When the decay regime is changed, the particle has a Lorentz factor $C_{curv}\sim10^4$ \eqref{Ccurv} and the oscillation amplitude is $A_{curv}\sim1$~m. Subsequently, the ultrarelativistic particle oscillations continue to decay exponentially and the oscillations become nonrelativistic and harmonic with a frequency $\nu\sim1-10$~GHz when an amplitude $l_{nro}\sim1$~cm is reached. Simultaneously with the oscillatory motion, the particle executes a regular motion over the force-free surface (see Fig. 3); the trajectory of the guiding center in the rotating frame of reference is generally closed and its velocity is of the order of the drift velocity.

Let us estimate the accumulation rate of an electron–positron plasma on the force-free surface. For this purpose, we considered a situation where the charged particles, electrons and positrons, moved in a given electromagnetic field while acting as test charges. This may be assumed only as long as the charge density in the magnetosphere is much lower than the Goldreich–Julian density $\rho_{GJ}=-\mathbf{B}\cdot\mathbf{\Omega}/2\pi c$. However, since the pair production rate in the magnetosphere is constant, particles increasingly accumulate with time on the force-free surface. Consequently, regions will appear near the force-free surface in the vacuum magnetosphere in some finite time in which the charge density will now be comparable to $\rho_{GJ}$. In the long run, the entire magnetosphere will be filled with plasma.

Let us investigate the initial stage of the rearrangement of the magnetosphere from a vacuum state to a state filled with plasma. Let us choose an arbitrary point $\mathbf{r}_0$ of the force-free surface and consider some small area on this surface containing the point in question. Since the sizes of the area are assumed to be small compared to $R$, the area itself is virtually flat. Charged particles accumulate with time on this area to form a symmetric charge layer. We will consider the plasma accumulation only on fairly short time scales, when the thickness of the forming charge layer is much smaller than $R$. In this case, the plasma density differs noticeably from zero only near the force-free surface; therefore, there is as yet no global rearrangement of the magnetosphere. For this reason, Eqs. \eqref{deutschMagneticField} and \eqref{deutschElectricField} specify the external electromagnetic field in which the charge layer is located. Let us find the intrinsic electric field of the layer at some distance $z$ from the force-free surface:
\begin{equation}
\label{intrinsicE}
E_l=4\pi\alpha\int\limits_0^z\rho_e(z')\,dz',
\end{equation}
where $\rho_e(z')$ is the volume charge density at distance $z'$ from the force-free surface and the distance $z$ does not exceed $h$---the distance from the force-free surface to the layer boundary. Here, it is important that the charge layer is symmetric, i.e., $\rho_e(-z')=\rho_e(z')$. To ensure an equilibrium state of this layer, it is necessary that each charged particle belonging to the layer undergo no acceleration along the magnetic field lines. In other words, the external longitudinal electric field must be completely compensated for by the projection of the intrinsic electric field of the layer \eqref{intrinsicE} onto the magnetic field direction. If we introduce the angle $\psi$ between the normal to the force-free surface and the magnetic field direction, then
\begin{equation*}
\label{cosPhi}
\cos\psi=\frac{\mathbf{B}\cdot\nabla(\mathbf{E}\cdot\mathbf{B})}
{B\,|\nabla(\mathbf{E}\cdot\mathbf{B})|}.
\end{equation*}
For an arbitrary $l$, the following relation then holds:
\begin{equation}
\label{zeroParallelFieldCondition}
4\pi\alpha\cos\psi\int\limits_0^{l\cos\psi}\rho_e(z')\,dz'=\omega^2l.
\end{equation}
Differentiating both parts of this equality with respect to $l$ yields $\rho_e=\omega^2/4\pi\alpha\cos^2\psi$. Thus, the charge density $\rho_e$ in the layer does not depend on the distance from the force-free surface at all. Recall that we measure the charge density in units of $e/{^-\!\!\!\!\lambda}^3$. It is determined only by the square of the frequency of nonrelativistic oscillations $\omega^2$ and the angle $\psi$, whose values are taken at point $\mathbf{r}_0$ around which we consider the small area of the force-free surface. The sign of the charge density $\rho_e$ coincides with that of $\omega^2$, in agreement with our conclusions after Eq.~\eqref{curvatureRadius}. In general, $\rho_e$ does not coincide with $\rho_{GJ}$. This stems from the fact that if the charge density in the layer were equal to the Goldreich–Julian density everywhere, then all particles in the layer would undergo complete corotation and, hence, would be at rest in the rotating frame of reference. In the layer, only the longitudinal electric field is zero but its transverse component is nonzero, because the intrinsic electric field of the layer does not necessarily lead to a nulling of the transverse component of the external field. For example, since the intrinsic electric field of the charge layer at point $\mathbf{r}_0$ is zero, the total electric field has only the transverse component, which causes the charged particles to drift over the force-free surface along the trajectories shown in Fig. 3. Hence, the forming charge layer is composed of differentially flowing currents on the force-free surface.

Thus, in the course of time, a charge layer whose thickness $h$ is a function of point $\mathbf{r}_0$ and time $t$ elapsed since the beginning of magnetosphere filling with plasma is formed near the force-free surface. It is important to note that only the charges of the same sign as that of the density $\rho_e$ itself contribute to the
charge density $\rho_e$ at each point of the force-free surface. Therefore, the charges in the layer are completely separated at the initial filling stage in accordance with the conclusions after Eq.~\eqref{curvatureRadius}. Consequently, the particle number density $n_e$ is equal in magnitude to the plasma density: $n_e=|\rho_e|$. At each point of the force-free surface, let us introduce the particle current density $j=dN/dSdt$, which coincides with the magnitude of the electric current density in dimensionless units. The equation for determining the thickness of the charge layer is then $\partial h/\partial t=j/n_e$. It can be used to estimate the time it takes for the entire vacuum magnetosphere to be filled with an electron–positron plasma. When the thickness of the charge layer $h$ becomes comparable to the characteristic size of the inner magnetosphere, i.e., the stellar radius $R$, a significant rearrangement of the entire magnetosphere will take place. The filling time $\tau_f$ is then $\tau_f=Rn_{GJ}/j$. Here, we took into account the fact that the plasma density in the layer is approximately equal to the Goldreich–Julian density, $n_e\approx n_{GJ}$. The flux density of the particles falling on the force-free surface is determined by the pair production mechanism in the magnetosphere. The gamma-ray photons producing electron–positron pairs in the magnetospheric magnetic field can fall into the magnetosphere from outside as a cosmic background and can be produced from soft thermal photons emitted by the stellar surface by Compton-scattered energetic particles. The main question that arises here is how many pairs can be produced by one photon with an energy above 1~MeV. Since the electrons and positrons being produced are rapidly accelerated to energies $\gamma_0$ \eqref{gammaMax}, a chain multiplication of pairs takes place. The number of produced pairs per primary photon depends exponentially on the characteristic size of the inner magnetosphere, $\exp{\mu}$, where $\mu=R/\ell$. Here, $\ell$ is the total mean free path of an energetic particle with respect to the generation of a curvature photon and its mean free path with respect to the production of a pair. The mean free path $\ell$ is known from an investigation of the stationary plasma generation processes in a polar magnetosphere \citep{GurevichIstomin1985} to be $\sim100$~m. This means that the multiplication factor can reach enormous values, of the order of $\exp{100}\simeq 10^{43}$. However, since there is an exponential factor, this question in the case of a vacuum magnetosphere requires a separate careful analysis, which is beyond the scope of this paper. A large pair multiplication factor can compensate for the smallness of the cosmic background photon flux. According to observations \citep{StrongEtal2005,SizunEtal2006}, the diffuse background $j_{ph}$ of Galactic photons with energies above 1~MeV is $\sim10^{-3}\text{ cm}^{-2}\text{s}^{-1}$. Therefore, the ratio $j_{ph}/cn_{GJ}\approx 10^{-25}$ is very small, but it does not prevent the pulsar magnetosphere from being rapidly filled with an electron–positron plasma at a sufficiently large pair multiplication factor.

\section*{Acknowledgements}

This work was supported by the Russian Foundation for Basic Research (project no. 08-02-00749-a) and a grant from the President of Russia for Support of Leading Scientific Schools (no NSh-1738.2008.2).

\begin{flushright}
\textit{Translated by V. Astakhov}
\end{flushright}
\newpage
\begin{figure}\centering
\includegraphics[height=0.85\textheight]{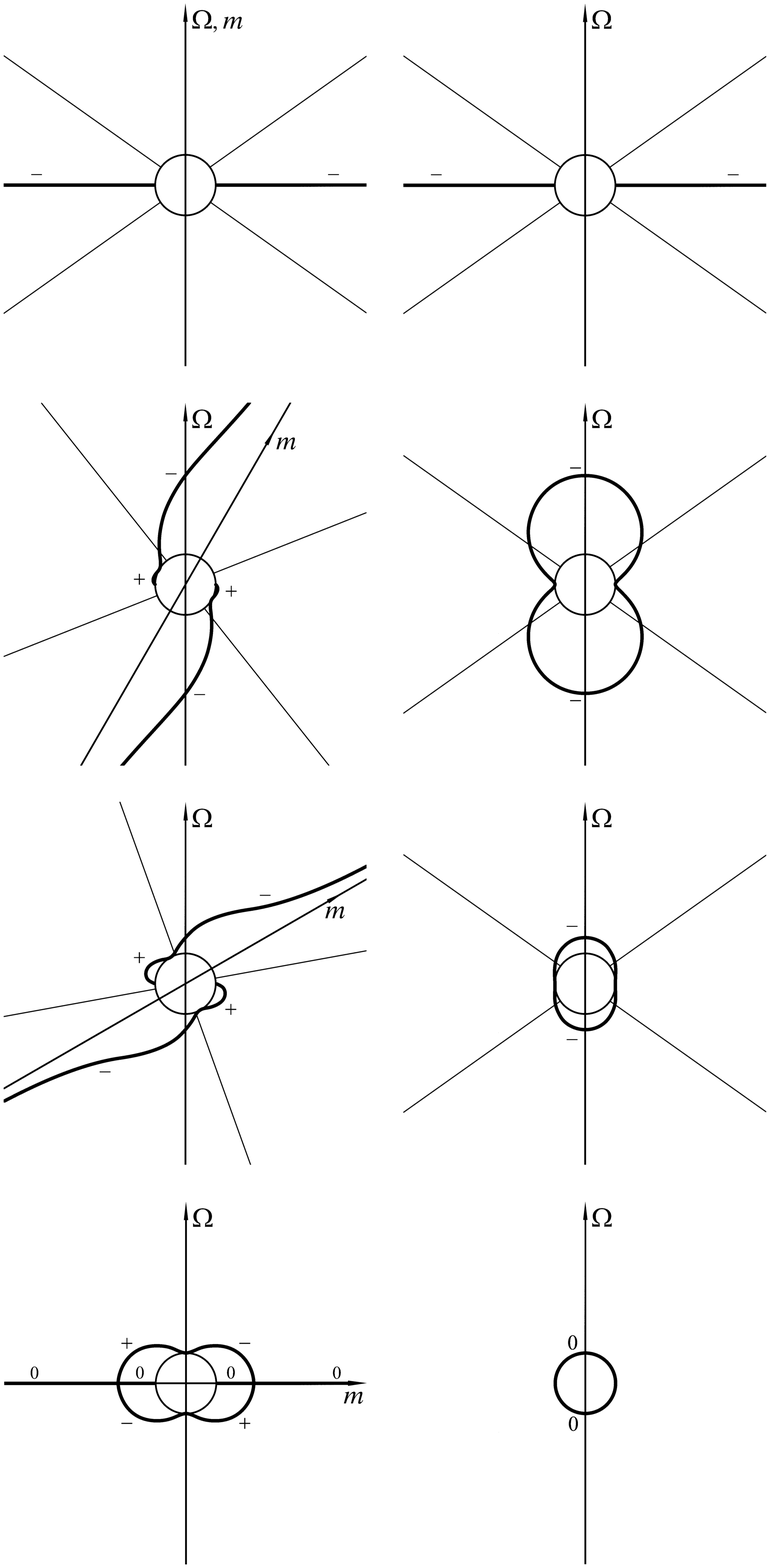}
\caption{Sections of the force-free surface by the planes $\varphi-\varphi_m=\{0,\pi\}$ (left) and $\varphi-\varphi_m=\{\pi/2,3\pi/2\}$ (right) for inclinations (from the top downward) of $0$, $\pi/6$, $\pi/3$, and~$\pi/2$.  The signs of the accumulating charges are marked. The sections $\rho_{GJ}=0$ are also shown for comparison.}
\end{figure}
\newpage
\begin{figure}
\includegraphics[width=\textwidth]{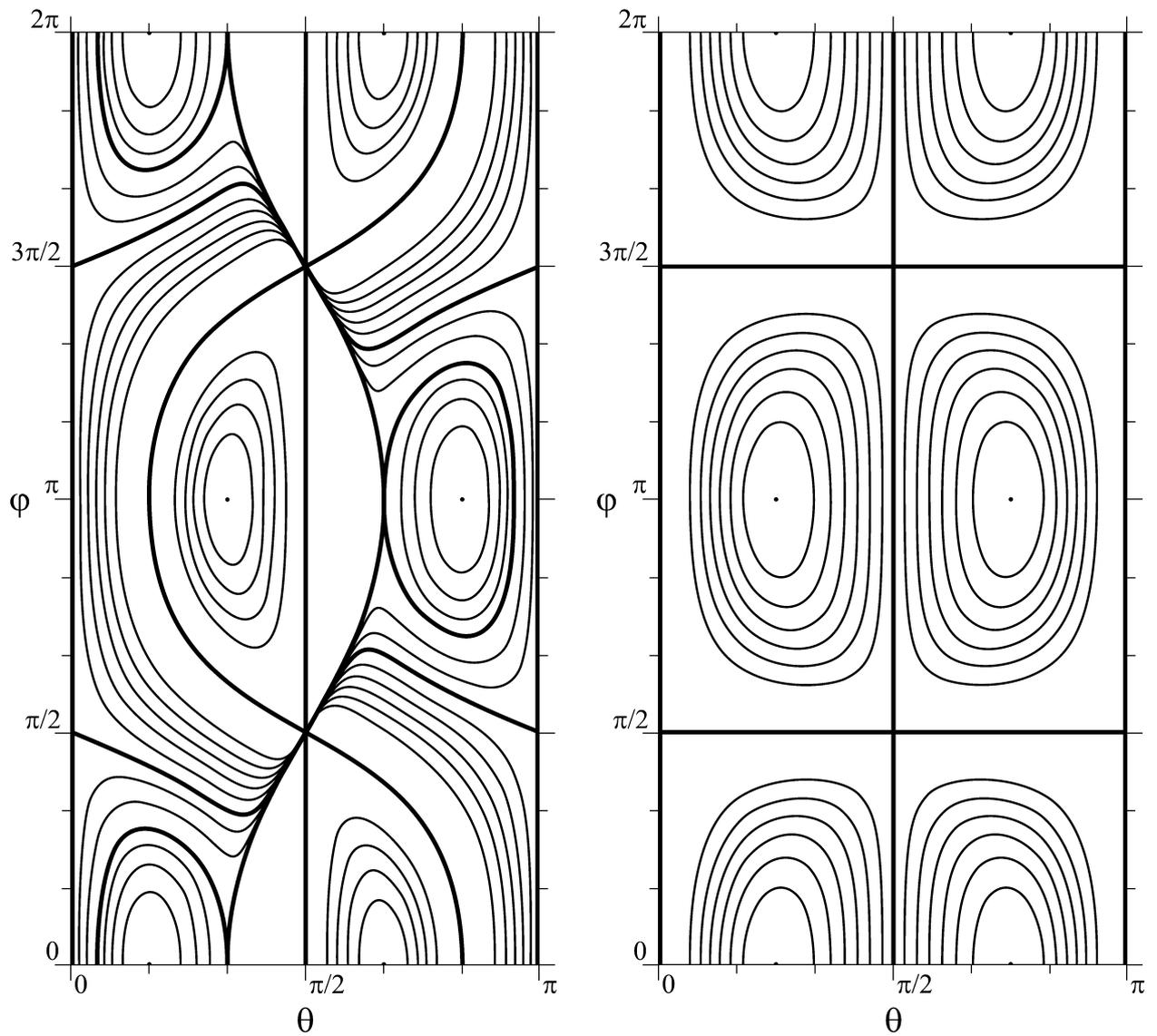}
\caption{Phase portraits of the charged particle trajectories on the force-free surface in $\theta$ and~$\varphi$ coordinates in the rotating frame of reference for an oblique rotator with an inclination of $\pi/3$ (left) and an orthogonal rotator (right).}
\end{figure}
\newpage
\begin{figure}
\includegraphics[width=\textwidth]{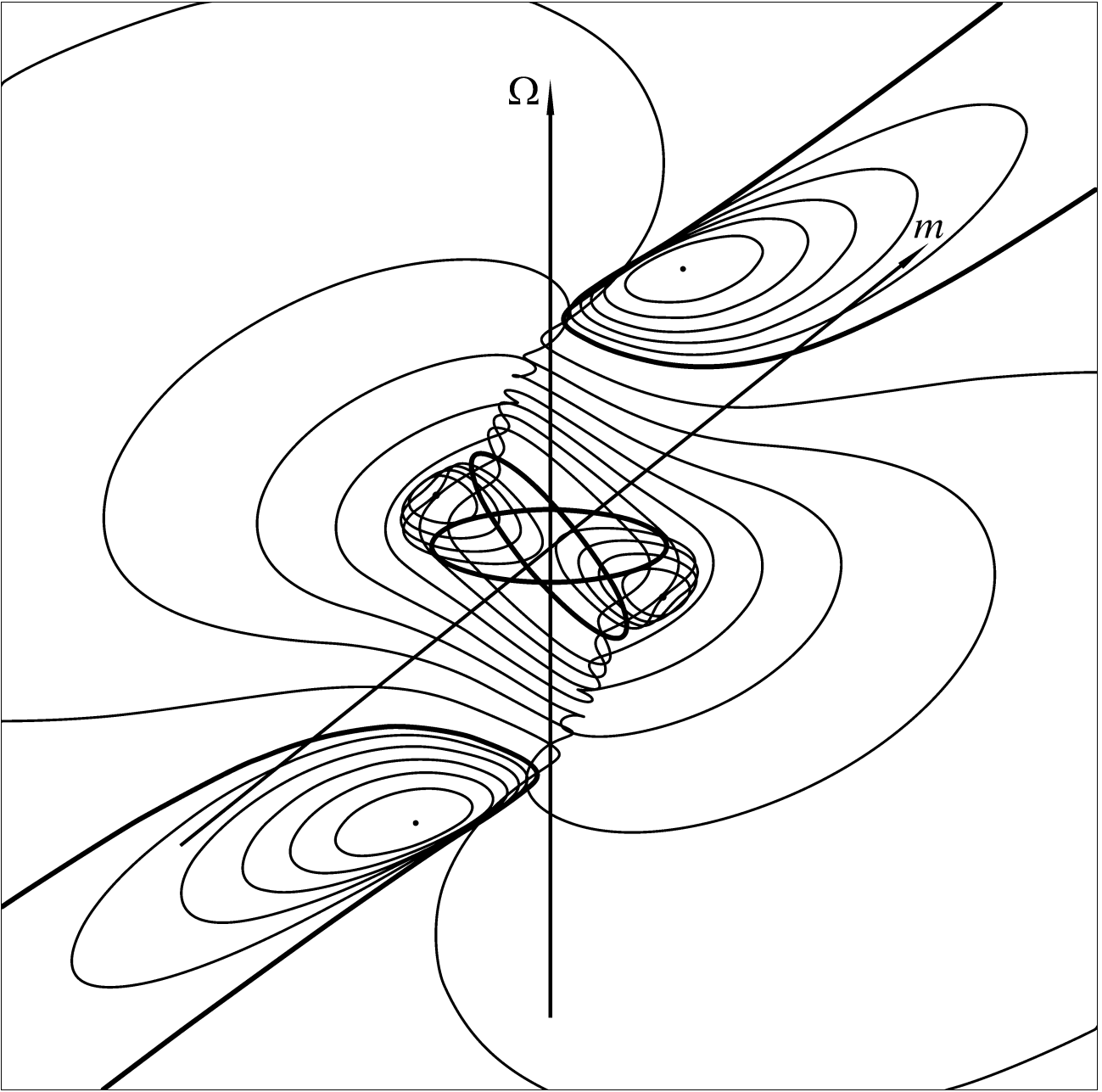}
\caption{Charged particle trajectories on the force-free surface in the rotating frame of reference for an oblique rotator with an inclination of $\pi/3$. The heavy lines indicate the equator, the magnetic equator, and the separatrices.}
\end{figure}
\end{document}